\documentclass[aip,prb,twocolumn,reprint,showpacs,superscriptaddress]{revtex4}
\usepackage{graphicx}
\usepackage{epsfig}
\usepackage{color}
\usepackage{amsmath}

\def\dd{\mbox{d}}

\def\kp{{k_{\rm{p}}}}
\def\ka{{k_{\rm{a}}}}
\def\drmp{{d_{\rm{p}}}}
\def\drma{{d_{\rm{a}}}}
\def\mua{{\mu_{\rm{a}}}}
\def\mup{{\mu_{\rm{p}}}}

\newcommand{\vgap}{\vspace{0.1cm}}

\begin{document}

\title{Enhancement of cargo processivity by cooperating molecular motors}

\author{Filippo Posta}
\affiliation{Dept. of Biomathematics, UCLA CA 90095-1766}
\email[]{fposta@ucla.edu}

\author{Maria R. D'Orsogna}
\affiliation{Dept. of Mathematics, CalState-Northridge, CA 91330-8313}
\email[]{dorsogna@csun.edu}

\author{Tom Chou}
\affiliation{Dept. of Biomathematics, UCLA CA 90095-1766}
\affiliation{Dept. of Mathematics, UCLA CA 90095-1766}
\email[]{tomchou@ucla.edu}

\date{\today}

\begin{abstract}
Cellular cargo can be bound to cytoskeletal filaments by one or
multiple active or passive molecular motors.  Recent experiments have
shown that the presence of auxiliary, nondriving motors, results in an
enhanced processivity of the cargo, compared to the case of a single
active motor alone.  We model the observed cooperative transport
process using a stochastic model that describes the dynamics of two
molecular motors, an active one that moves cargo unidirectionally
along a filament track and a passive one that acts as a tether.
Analytical expressions obtained from our analysis are fit to
experimental data to estimate the microscopic kinetic parameters of
our model. Our analysis reveals two qualitatively distinct
processivity-enhancing mechanisms: the passive tether can decrease the
typical detachment rate of the active motor from the filament track or
it can increase the corresponding reattachment rate. Our estimates
unambiguously show that in the case of microtubular transport, a
higher average run length arises mainly from the ability of the
passive motor to keep the cargo close to the filament, enhancing the
reattachment rate of an active kinesin motor that has recently
detached. Instead, for myosin-driven transport along actin, the
passive motor tightly tethers the cargo to the filament, suppressing
the detachment rate of the active myosin.  
\end{abstract}


\maketitle


\section{Introduction}
\label{intro}
To carry out its functions, a living cell requires the precise
spatiotemporal organization of many macromolecules.  Trafficking of
molecules within the cytoplasm can be mediated by distinct processes
including diffusion, polymerization, and active transport
\cite{Vale03}. A variety of transport mechanisms may arise from the
physical properties of the diverse cargoes being transported. For
example, in the case of large cargoes, such as organelles, mRNA or
virus particles, diffusion may not be sufficiently fast nor be spatially
controlled.  These cargoes are often transported by motor proteins that
processively move to and from the nucleus along specific cytoskeletal
filaments \cite{Mallik06}.

The cytoskeleton is typically composed of three types of filaments:
microfilaments ({\it e.g.} actin), microtubules ({\it e.g.} tubulin
$\alpha$ and $\beta$), and intermediate filaments ({\it e.g.}  lamins)
\cite{Lodish05}. Molecular motors most often associate with and
process cargo along actin and microtubules \cite{Gross07review}. These
two filament types are structurally very different from each other.
Microtubules (MT) are thicker (25nm diameter) and have a specific
radial orientation with respect to the cell nucleus. Actin filaments
are more randomly distributed near the periphery of the cell, and are
less thick (8nm diameter) than MTs \cite{Mallik06,Lodish05}.
Moreover, filaments are directional. The ends of a microtubule are
structurally different and labelled ``positive'' or ``negative,''
while the ends of actin filaments are ``pointed'' or ``barbed.''
Accordingly, there are different types of motor proteins associated
with not only different filament types, but with the direction of
transport along these filaments.  The MT-specific motor proteins are
kinesins ({\it e.g.}, kinesin I, II) that can move along MTs away from
the nucleus on the positive direction, and dyneins that move in the
opposite direction towards the negative end of a microtubule
\cite{Welte04}.  Various forms of the actin-specific motor myosin
transport cargo toward the barbed ({\it e.g.}  myoV) or pointed ({\it
e.g.} myoVI) ends \cite{Metha99,Gross07}. Since each motor is highly
selective, and cargoes need to be moved on both directions of each
filament, there are other proteins/cofactors that facilitate molecular
transport by associating with specific motors and filaments. For
example, dynactin is a cofactor of dynein that enhances both the
processivity of dynein and its affinity to certain cargoes
\cite{Welte04}.

Single molecule imaging methods have been pivotal in the experimental
study of molecular motor dynamics \cite{Block90,Visscher99}. Such
advanced techniques have allowed researchers to dissect various
aspects of molecular motor mediated transport
\cite{Shiroguchi07,Rief00,Carter05,Svoboda94,Kawaguchi00,Clemen05}.
The identification of motor proteins, their structure and properties
have also led to several theoretical studies
\cite{Fisher99,Kolomeisky03,Skau06,Kolomeisky07} that have further
improved our understanding of how a single motor protein is able to
move a cargo along a cytoskeletal filament.

Since many associated proteins/cofactors affect transport dynamics
\cite{Gross07review}, experimental and theoretical investigations of
model systems that include only one motor and the tracks on which they
bind do not yield a complete description of molecular motor based
transport {\it in vivo}.  Therefore, other recent studies have focused
on how cooperativity among different molecular motors can facilitate
cargo transport along straight, branched, and intersecting
cytoskeletal filaments \cite{Gross07,Vilfan08,Kunwar08}. In the
experiments of Ali {\it et al.} \cite{Ali07,Ali08} the cargoes were
fluorescently labeled quantum dots (Qdots) \cite{Alivisatos05}
simultaneously attached to one kinesin and one myoV motor.  While
analyzing the dynamics of myoV transport in the presence of both actin
and MT filaments, the authors discovered that myoV, besides processing
along actin, is also able to associate with, and diffuse along
microtubules. In further work \cite{Ali08}, the same authors showed
that when both myoV and kinesin motors are attached to a single cargo,
the processivity of the entire assembly is increased on both MT and
actin filaments.

\begin{figure}[tb]
\begin{center}
\includegraphics[width=8cm]{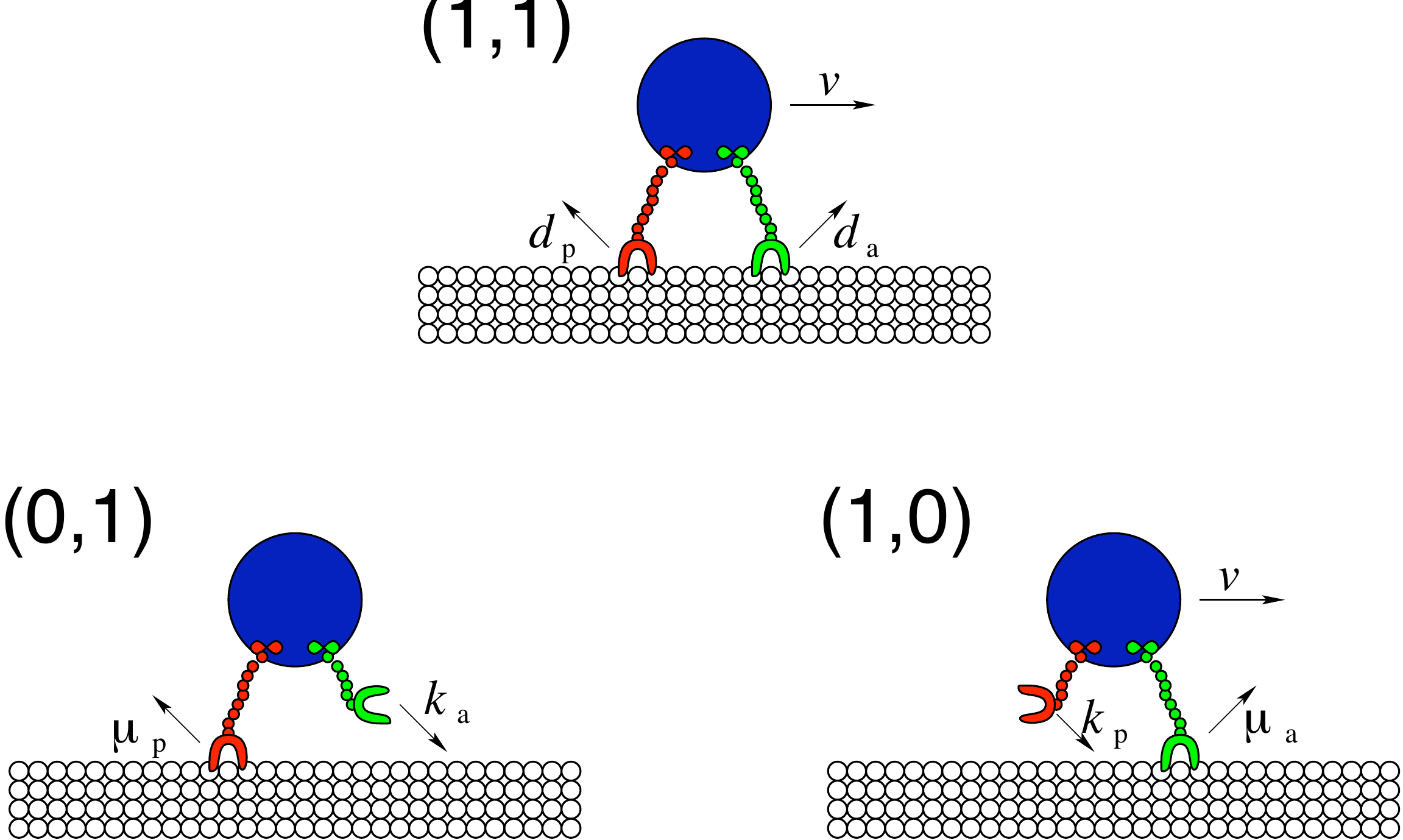}
\end{center}
\caption{Schematic of movement along a filament track of a cargo (blue
circle) with both an active (green) and passive (red) motors
attached. Each figure represents one of three possible states the
cargo complex can be in before detachment, together with the rate
constants that lead the system out of a state. Note that the complex
moves processively only if the active motor is attached. Only in state
$(0,1)$ can the system diffuse.}
\label{fig:FIG1}
\end{figure}

In this paper, we develop a stochastic model for the cooperative
enhancement of kinesin and myoV as discovered in \cite{Ali08}. Related
stochastic models, both discrete and continuous have been effectively
used to model single motor dynamics \cite{Skau06,Milescu06} and
cooperativity \cite{Klumpp05}.  When applied to single motors, these
stochastic models describe the stepping dynamics of typically
two-headed molecular motors ({\it e.g.}  myoV or kinesin) that walk
hand-over-hand along the filament, where the trailing and the leading
heads exchange roles at each step.  In our analysis, we take a coarser
approach by treating the entire kinesin-myoV-cargo complex as a single
structure that can exist in four possible states corresponding to
different association states of each motor with each filament type.
From the model we extract and briefly discuss the expressions for the
mean run length and the first passage time before detachment. In the
Analysis \& Discussion section, we fit the model to the experimental
results from \cite{Ali07, Ali08}. The fitting highlights a qualitative
difference between cargo transport on actin and microtubules. We also
discuss the dependence upon the initial conditions of the system and
perform a sensitivity analysis on the unknown parameters.\\


\section{Stochastic Multistate Transport  Model}
\label{sec:model}
Consider the molecular transport system of a cargo with two
motor-proteins attached, one that acts as an active motor ({\it e.g.}
kinesin on MT) and one that act as a passive motor or tether ({\it
e.g.} myoV on MT).  We denote the state of the engine $\sigma_a$ and
of the passive tether $\sigma_p$ by an ordered pair $\sigma =
(\sigma_a, \sigma_p)$, where $\sigma_{a,p} = 1$ if the active/passive
motor is attached to the filament track and where $\sigma_{a,p} = 0$
otherwise. Using this notation, state $\sigma=(1,1)$ corresponds to
the case where both engine and tether attach cargo to the
filament. State $(1,0)$ represents a cargo complex whose engine is
attached, but where the tether has detached from the filament
(although it remains attached to the cargo).  Conversely, $(0,1)$
denotes the case where the engine has detached from the filament, but
the tether still holds the cargo on the track.  Finally, when both
motors have detached from the filament, the complex reaches the
$(0,0)$ state. The states of the cargo system are schematically shown
in Fig. \ref{fig:FIG1}. We only consider states in which, at
any given time, at most only one active motor and one passive motor
tether can attach a cargo to a filament. This assumption is reasonable
because the procedure used in \cite{Ali08} for motor attachment
predicts that 95$\%$ of the Qdots used as cargo have only one active
motor attached. While there are no predictions about the number of
tethers attached to the cargo, the relatively small size of the Qdots
used in the experiments ($\sim15$ nm diameter) suggests that
simultaneous attachment of an active motor and multiple tethers is
highly unlikely.

The cargo complex can move processively along a filament only if the
active motor is attached to both cargo and filament. In our notation,
this corresponds to states with first index one, {\it e.g.} $(1,1)$ or
$(1,0)$. In state $(0,1)$, the cargo is either diffusing or immobile,
depending on the property of the passive motor.  Within this context,
we can write the Master equation for the probability density function
$P_{\sigma} (x,t)$ that the motor-cargo complex is in state $\sigma$
between position $x$ and $x+\dd x$ at time $t$:

\begin{widetext}
\begin{eqnarray}
\label{eqn:master}
{\partial P_{\sigma}(x,t) \over \partial t} + v_{\sigma}{\partial 
P_{\sigma}(x,t)\over \partial x} &=& D_{\sigma}{\partial^{2}P_{\sigma}
(x,t) \over \partial x^{2}} + \sum_{\sigma'} \left [W(\sigma,\sigma') 
P_{\sigma'}(x,t)- W(\sigma',\sigma) P_{\sigma}(x,t) \right].
\end{eqnarray}
\end{widetext}

\noindent Here, $v_{\sigma}$ is the velocity of the cargo in state
$\sigma$.  This velocity will depend on the specific properties of the
motor-protein and on any externally applied forces. For example, as
has been well established under many experimental conditions
\cite{Kawaguchi00,Kunwar08} opposing forces applied to motor-driven
cargoes linearly decrease their velocity.

Consistent with observations, we set the diffusion constant
$D_{\sigma}=0$ in Eq. \ref{eqn:master} when an active, driving motor
is attached, suppressing random diffusional motion. Conversely, when
only a passive tether is attached, the motion of the cargo is Brownian
and $D_{\sigma} > 0$.

The last term in Eq. \ref{eqn:master} represents transitions among
binding states $\sigma$. The corresponding rates $W(\sigma, \sigma')$ are
assumed to be constant and are defined in Table
\ref{tab:transition}. Note that we make the physically reasonable
assumption that a cargo complex in state $(1,1)$ cannot have both
motors detach simultaneously from the filament track, allowing no
transition between state $(1,1)$ and $(0,0)$.

We will analyze the model given by Eq. \ref{eqn:master}  by defining the
probability density vector ${\bf P}(x,t) = 
(P_{(0,0)},P_{(0,1)},P_{(1,0)},P_{(1,1)})^{T}$.
%
%
If we assume that the filament track is infinitely long and that the
cargo is at position $x=0$ at initial time $t=0$, the initial
condition is ${\bf
P}(x,t=0)=(0,\alpha,\beta,1-\alpha-\beta)^T\delta(x)$, where $\alpha$
is the probability that the cargo complex is initially bound to the
filament only by the passive tether, and $\beta$ is the probability
that the cargo complex is initially bound only to the active motor.
While it is experimentally difficult to quantify $\alpha$ and $\beta$,
it is relatively straightforward to determine how our main results for
residence times and run lengths depend on the initial conditions.  We
can therefore establish, {\it a posteriori}, the importance of
$\alpha, \beta$ in the measured results.  Thus, by studying how
certain estimated quantities depend on the initial conditions, we can
determine the significance and usefulness of experimentally
pinpointing the exact values of $\alpha$ and $\beta$.

\begin{table}[tb]
\begin{center}
\caption{Description of transition rates $W(\sigma, \sigma')$ in Eq. \ref{eqn:master}.}
\label{tab:transition}
\begin{tabular}{|l|c|c|}
\hline
State Transition & Rate & Description \\
\hline
$(1,1) \rightarrow (0,1)$ & $d_{\rm{a}}$ & \parbox[c]{2in}{\vgap Detachment 
rate of motor from motor-tether complex.\vgap}\\
$(1,1) \rightarrow (1,0)$ & $d_{\rm{p}}$ & \parbox[c]{2in}{\vgap Detachment 
rate of tether from motor-tether complex.\vgap}\\ 
$(0,1) \rightarrow (1,1)$ & $k_{\rm{a}}$ & \parbox[c]{2in}{\vgap Attachment 
rate of motor to tether only complex.\vgap}\\
$(1,0) \rightarrow (1,1)$ & $k_{\rm{p}}$ & \parbox[c]{2in}{\vgap Attachment 
rate of tether to motor only complex.\vgap}\\
$(1,0) \rightarrow (0,0)$ & $\mu_{\rm{a}}$ & \parbox[c]{2in}{\vgap Detachment 
rate of motor from motor only complex.\vgap}\\
$(0,1) \rightarrow (0,0)$ & $\mu_{\rm{p}}$ &  \parbox[c]{2in}{\vgap Detachment 
rate of tether from tether only complex.\vgap}\\
\hline
\end{tabular}
\end{center}
\end{table}

The analysis is facilitated by defining the Laplace transform in time
$\tilde{{\bf P}}(x,s) = \int_{0}^{\infty}{\bf P}(x,t) e^{-st} \dd t$,
and by taking the dual Laplace-Fourier transform of
Eq. \ref{eqn:master}:

\begin{widetext}
\begin{equation}\label{eqn:metransform}
s\hat{{\bf P}}(q,s) - \left(\begin{array}{c}
0\\[13pt]
\alpha \\[13pt]
\beta \\[13pt]
1-\alpha-\beta \end{array}\right) + \left(\begin{array}{cccc} 0 & 0 & 0 & 0 \\[13pt]
0 & D_p \; q^{2} & 0 & 0 \\[13pt]
0 & 0 & iqv_{a} & 0 \\[13pt]
0 & 0 & 0 & iqv_{a} \end{array}\right) \hat{{\bf P}}
=  \left(\begin{array}{cccc} 0 & \mu_{\rm{p}} & \mu_{\rm{a}} & 0 \\[13pt]
0 & -k_{\rm{a}}-\mu_{\rm{p}} & 0 & d_{\rm{a}} \\[13pt]
0 & 0 & -k_{\rm{p}}-\mu_{\rm{a}} & d_{\rm{p}} \\[13pt]
0 & k_{\rm{a}} & k_{\rm{p}} &  -d_{\rm{a}}-d_{\rm{p}} \end{array}\right) 
\hat{{\bf P}},
\end{equation}
\end{widetext}
\noindent
where $\hat{{\bf P}}(q,s) =\int_{-\infty}^{\infty}\tilde{{\bf
P}}(x,s)e^{-iqx}\dd x$.  In the above equation, and in the remainder
of the paper, we use the subscripts $a$, $p$ and $ap$ to indicate the
quantities and/or expressions that are characteristic of a transport
complex consisting of an active motor only ($a$), a passive motor only
($p$) or both ($ap$).  In Eq. \ref{eqn:metransform}, we assumed that
motion is purely convective when an active motor attaches the cargo to
the filament, and therefore set $D_a=D_{ap}=0$.  As indicated by the
experimental data in \cite{Ali08}, the passive tether does not
noticeably affect the transport velocity of an active
motor. Therefore, we also set $v_{ap}=v_{a}$, where $v_{ap}$ is the
intrinsic velocity of cargo in the state $(1,1)$.  Finally, since the
passive motor acting as a simple tether cannot induce drift along a
filament track, we set $v_{p}=0$. After imposing these physical
constraints, we solve the Master equation to obtain analytical
expressions for the mean run length and time. We will use these
expressions to compare our model with experimental data from
\cite{Ali08}.

\subsection{Mean Run Lengths}
\label{sec:runlength}

Having more than one motor attached to a cargo complex typically
results in an improved transport efficiency.  Experiments by Ali {\it
et al.}  \cite{Ali08} show significant increases in the processivity
of a cargo when it is also attached to a {\it passive} motor. Within
our model, the measured processivity is equivalent to the mean run
length, $\langle X_{ap} \rangle$, of the cargo complex before
detachment.  To find its expression, we construct the detachment flux
density:

\begin{equation}\label{eqn:jap}
J_{ap}(x,t) = \mu_{\rm{a}}P_{(1,0)}(x,t) + \mu_{\rm{p}}P_{(0,1)}(x,t).
\end{equation}
We can use the properties of the Laplace and Fourier transforms to
find a compact expression for $\langle X^m_{ap}\rangle$, the $m^{\rm th}$
moment of the run length before detachment:

\begin{equation}\label{eqn:rlmoments}
\begin{array}{rl}
 &  \langle X_{ap}^{m}\rangle \displaystyle 
 = \int_{0}^{\infty}\left[\int_{0}^{\infty} J_{ap}(x,t) \dd t\right] x^{m} \dd x \\[13pt]
 &  \displaystyle = \int_{0}^{\infty} x^{m} \left[\mu_{\rm{a}}\tilde{P}_{(1,0)}(x,s=0) 
+ \mu_{\rm{p}}\tilde{P}_{(0,1)}(x,s=0)\right]\dd x \\[13pt]
 &  \displaystyle = \left(i{\partial \over \partial q}\right)^{m} 
\hat{J}_{ap}(q,s=0)\bigg|_{q=0}.
\end{array}
\end{equation}
\begin{table}
\begin{center}
\caption{Typical values of parameters and mathematical quantities for 
kinesin/myoV cargo transport.}
\label{tab:parametervalues}
\begin{tabular}{|l|c|c|r|}
\hline
 & Mean Value & Biophysical Setup & Ref. \\
\hline 
$\langle X_{a} \rangle$ & $ \left\{ \begin{array}{c}0.76 \; \mu m \\ 1.7 \; \mu m \end{array} 
\right.$ & $\begin{array}{c} \mbox{myoV on actin} \\ \mbox{ 
kinesin on MT} \end{array}$  & \cite{Ali08}\\[13pt]
$\langle X_{ap} \rangle$ & $ \left\{ \begin{array}{c} 1.09 \; \mu m \\ 3.7 \; \mu m 
\end{array} \right.$ & $\begin{array}{c} \mbox{myoV-kinesin on actin} \\ 
\mbox{myoV-kinesin on MT} \end{array}$  & \cite{Ali08}\\[13pt]
$v_a$ & $ \left\{ \begin{array}{c}0.46 \; \mu m/s \\ 0.88 \; \mu m/s \end{array} 
\right.$ & $\begin{array}{c} \mbox{myoV on actin} \\ \mbox{ 
 kinesin on MT} \end{array}$  & \cite{Ali08}\\[13pt]
$\mu_a$ & $ \left\{\begin{array}{c} 0.60 \; s^{-1} \\ 0.51 \; s^{-1} \end{array} 
\right.$ & $\begin{array}{c} \mbox{myoV from actin} \\ \mbox{ 
kinesin from MT} \end{array}$  & \cite{Ali08}\\[13pt]
$k_a$ & $>0.2 \; s^{-1}$ & kinesin to MT & \cite{Ali08} \\[8pt]
$\mu_p$ & $>0.02 \; s^{-1}$ & myoV from MT 
& \cite{Ali07} \\[8pt]
$D$ & $0.11-0.26 \; \mu m^2/s$ & myoV on MT&
\cite{Ali07,Ali08} \\[5pt]
\hline
\end{tabular}
\end{center}
\end{table}

\noindent
By taking $m=1$ we can find the expression for the mean run length of the
motor-tether complex before detachment

\begin{widetext}
\begin{eqnarray}\label{eqn:Xap}
\langle X_{ap} \rangle & = &v_{a}\frac{
\left(\drmp + \kp + \mua - \beta \mua
\right) \left(\ka + \mup \right) 
+  \drma  \beta \mup - \alpha \left[ \mup
\left( \kp + \mua \right) +  \drmp \mup
\right]}{\drma \; \mup (\kp + \mua) + \drmp \; \mua (\ka + \mup)}.
\end{eqnarray}
\end{widetext}
As expected, Eq. \ref{eqn:Xap} is independent of any diffusion of the
passive tether since diffusion on average does not contribute to the
mean displacement. The mean run length is a monotonically decreasing
function of $\alpha$ for all the physically realistic ({\it i.e.}
positive) values of the model parameters.  On the other hand, the
dependence of $\langle X_{ap} \rangle$ on $\beta$ will be
monotonically increasing (decreasing) if the term $\drma \mup - \mua
\left(\ka + \mup \right)$ is positive (negative).
The processivity of a cargo complex initially in state $(1,0)$ (when
$\beta =1$) is greater than that of a cargo beginning in state $(1,1)$
(when $\beta=0$) only when the detachment rate $d_{\rm a}$ of the
active motor from the state when both motors are attached is greater
than $\mua(1 + \ka/\mup)$, where $\mua$ is the detachment rate of the
active motor by itself.

Establishing the dependency of Eq. \ref{eqn:Xap} upon the remaining
parameters does not lead to an easy expression unless we make some
simplifications.  We can focus on the limit in which all cargo
complexes have both motors attached to the track ($\alpha=\beta=0$) at
time zero. This is also the underlying assumption used in
\cite{Ali08}. In this case, we find that the mean run length before
detachment reduces to

\begin{equation}\label{eqn:Xapsimple}
\langle X_{ap} \rangle = \frac{(k_{\rm{p}}+
d_{\rm{p}}+\mu_{\rm{a}})(k_{\rm{a}}+\mu_{\rm{p}})v_{a}} 
{\mu_{\rm{p}}d_{\rm{a}}(k_{\rm{p}}+\mu_{\rm{a}}) + 
\mu_{\rm{a}}d_{\rm{p}}(k_{\rm{a}}+\mu_{\rm{p}})}.
\end{equation}
As expected, $\langle X_{ap} \rangle$ is a monotonically increasing
function of $\ka$, $\kp$ and $v_a$, and a monotonically decreasing
function of $\drma$, $\drmp$, $\mua$ and $\mup$.

The probability density flux $J_{a}$ out of the state
where the driving motor is attached can be expressed as

\begin{equation}
J_{a}(x,t) = \mu_{\rm{a}}P_{(1,0)}(x,t) + d_{\rm{a}}P_{(1,1)}(x,t).
\end{equation}
Since the final two states $(0,\sigma_{p})$ are both absorbing, we
must also set $k_{\rm{a}} = 0$ in the evaluation of $J_{a}(x,t)$. The
expression for the moments of this density flux is analogous to
Eq. \ref{eqn:rlmoments}. The mean run length of a cargo complex
conditioned on being bound by an active motor is given by

\begin{eqnarray}\label{eqn:Xa}
\langle X_{a} \rangle & = & v_a \frac{ \beta \left( 
\drma - \mua \right) +  \left( \drmp + \kp + \mua 
\right) }{\drmp \mua + \drma \left( \kp + \mua \right)},
\end{eqnarray}
and is a monotonically increasing (decreasing) function of $\beta$ if
$\drma > \mua$ ($\drma < \mua$). Similarly, a simple analysis of
Eq.\,\ref{eqn:Xa} reveals that if $\beta=0$ the mean run length
$\langle X_a \rangle$ in state $(1,\sigma_{p})$ is a monotonically
decreasing function of $\kp$ and a monotonically increasing function
of $\drmp$ if $\drma > \mua$.  If $\drma < \mua$, the mean run length
$\langle X_a \rangle$ in state $(1,\sigma_{p})$ is a monotonically
increasing function of $\kp$ and a monotonically decreasing function
of $\drmp$. When the detachment rate of the active motor is
independent of its state ({\it e.g.}, $\mua=\drma$), $\langle X_{a}
\rangle = v_a/\mua$ is independent of the initial conditions and the
properties of the tether.

Finally, the mean run length of a cargo bound only by a passive tether
(state $(0,1)$) can be found from 

\begin{equation}\label{eqn:Jp}
J_{p}(x,t) = \left(\mup + \ka \right) P_{(0,1)}(x,t),
\end{equation}
and setting the rates into $(0,1)$ from all absorbing states
$\drma=\beta=0$, yielding  $\langle X_{p}\rangle = 0$. This trivial result stems
from the drift-free nature ($v_{p}=0$) of the $(0,1)$ state.

\begin{figure}[tb]
\begin{center}
  \includegraphics[width=8cm]{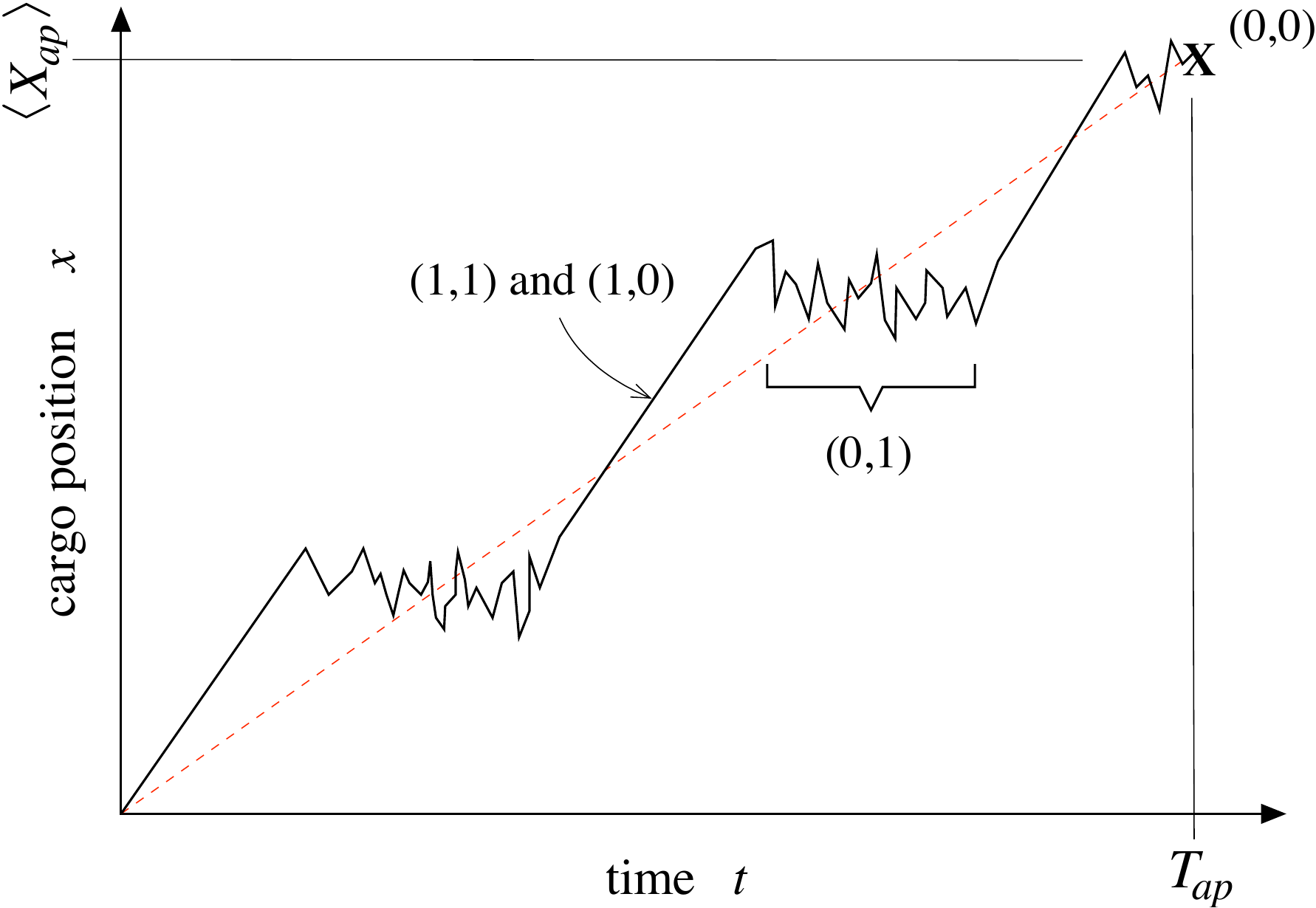}
\end{center} 
\caption{A representative trajectory. The processive runs during
states $(1,1)$, $(0,1)$ are assumed to have the same velocity
independently of the state they are in.  The depicted cargo movement
is characterized by three processive runs, all with the same slope
({\it e.g.} same velocity) and three diffusive events due to the cargo
complex being in state $(0,1)$ before ccomplete detachment at ${\bf
X}$.  The dashed red line represents a straight trajectory that
defines an effective velocity of a complete run before detachment.}
\label{fig:diffusiverun}
\end{figure}

\subsection{First Passage Times}
\label{sec:ftp}
Together with the mean run length, the data in \cite{Ali08} include
the average velocities for all experimental motor/tether
configurations. To use this data for parameter fitting, we derive
analytical expressions for the average duration of a run, {\it i.e.},
the first passage time to the detached state.  In order to evaluate
the detachment time of the driving motor anywhere along the filament
track, we consider the conditional mean first passage time $T_a$ out
of states $(1,\sigma_{p})$, for which $\alpha = 0$, as follows
 
\begin{eqnarray}\label{fpt}
T_{a} &=&  \frac{\int_{0}^{\infty}\dd t \int_{-\infty}^{\infty}\dd x \,\, t J_{a}(x,t)}{
  \int_{0}^{\infty}\dd t \int_{-\infty}^{\infty}\dd x  \,\, J_{a}(x,t)} \nonumber \\
  & = &\left.  - \frac{1}{\hat{J}_{a}(q=0,s)}
\frac{\partial \hat{J}_{a}(q=0,s)}{\partial s}\right|_{s=0}.
\end{eqnarray}

\noindent 
Upon using  Eq.\,\ref{fpt} and $\ka=0$ together with the initial
condition ${\bf P}(x,0) = (0, 0, \beta, 1 -\beta)^{T}\delta(x)$ we find

\begin{eqnarray}\label{eqn:Ta}
T_{a} & = & \frac {\beta (\drma - \mua) + (\drmp + \kp + \mua)}\nonumber \\
& = & {\drmp \mua + \drma (\kp + \mua)}.
\end{eqnarray}
We can use this result in conjunction with the expression for the mean run length from 
Eq. \ref{eqn:Xa} to obtain an estimate of the velocity $V_a$ of the cargo 
conditioned on it being attached by the active motor:

\begin{eqnarray}\label{eqn:Va}
V_a=\frac{\langle X_a \rangle}{T_{a}} = v_a .
\end{eqnarray}
Since the passive motor does not affect the motion of the cargo, the
velocity when an active motor is attached is independent of
the binding and unbinding of the passive tether. This velocity 
is that within a single processive run. The velocity averaged 
over a trajectory composed of both processive and diffusive runs 
will typically be smaller, as discussed below.


We now compute $T_{ap}$, the mean first time to detachment (state
$(0,0)$). Using analogous notation, we find

\begin{eqnarray}
T_{ap} &=& \frac{\int_{0}^{\infty}\dd t
\int_{-\infty}^{\infty}\dd x \,\, t J_{ap}(x,t)}{ \int_{0}^{\infty}\dd t
\int_{-\infty}^{\infty}\dd x \,\, J_{ap}(x,t)} \nonumber \\
&=& \left. -
\frac{1}{\hat{J}_{ap}(q=0,s)} \frac{\partial
\hat{J}_{ap}(q=0,s)}{\partial s} \right|_{s=0},
\end{eqnarray}

\noindent from which we obtain

\begin{widetext}
\begin{eqnarray}\label{eqn:Tap}
T_{ap} = \frac{(\drma + \ka) (\kp + \mua - 
    \beta \mua) + (\beta (\drma - \mua) + (1- \alpha) (\kp + \mua)) \mup + 
\drmp (\ka + \alpha (\mua - \mup) + \mup)}{
\drma (\kp + \mua) \mup + \drmp \mua (\ka + \mup)}.
\end{eqnarray}
\end{widetext}
As in section \ref{sec:runlength}, we can study the
dependence of $\langle T_{ap}\rangle$ on the kinetic parameters
by considering the $\alpha=\beta=0$ limit:

\begin{equation}\label{eqn:Tapsimple}
T_{ap}= \frac{\drmp (\ka + \mup) + (\kp + \mua) (\drma + \ka + \mup)}
{\drma (\kp + \mua) \mup + \drmp \mua (\ka + \mup)}.
\end{equation}

\noindent
This expression is monotonically decreasing with respect to $\mua$ and
$\mup$, and can be used in conjunction with Eq. \ref{eqn:Xapsimple} to
obtain an expression for the mean velocity of cargo transport in
presence of an active and a passive motor

\begin{equation}\label{eqn:Vapsimple}
\langle V_{ap} \rangle = \frac{(k_{\rm{p}}+
d_{\rm{p}}+\mu_{\rm{a}})(k_{\rm{a}}+\mu_{\rm{p}})v_{a}} 
{\drmp (\ka + \mup) + (\kp + \mua) (\drma + \ka + \mup)}  \; \le \; v_a.
\end{equation}

\noindent
This result indicates that although in states $(1,1)$ and $(1,0)$ the
cargo drifts with velocity $v_a$, the entire run consists of
alternating phases of drifting and diffusive states, resulting in an
average effective velocity $V_{ap} \le v_a$, with the two velocities being 
equal only under certain regimes such as for slow active motor dissociation from 
state $(1,1)$ ({\it e.g.} $\drma \ll 1$) or for very fast tether dissociation from 
state $(0,1)$ ({\it e.g.} $\mup \gg 1$). From the above expression one can 
estimate the ratio $\chi$ of convection times to diffusion times as follows

\begin{eqnarray}
\chi &=& \frac{|V_{ap} - V_{a}|}{v_a}=  \nonumber \\ 
&=& \frac{\drma (\kp + \mua)}{\drmp
(\ka + \mup) + (\kp + \mua) (\drma + \ka + \mup)} \; \le \; 1. 
\end{eqnarray}


The above algebraic expressions for the mean length ({\it e.g.} Eq.
\ref{eqn:Xap}) and duration ({\it e.g.} Eq. \ref{eqn:Tap}) of a run
out of the various states constitute our  main mathematical results.

\begin{figure}[tb]
\begin{center}
\includegraphics[width=9cm]{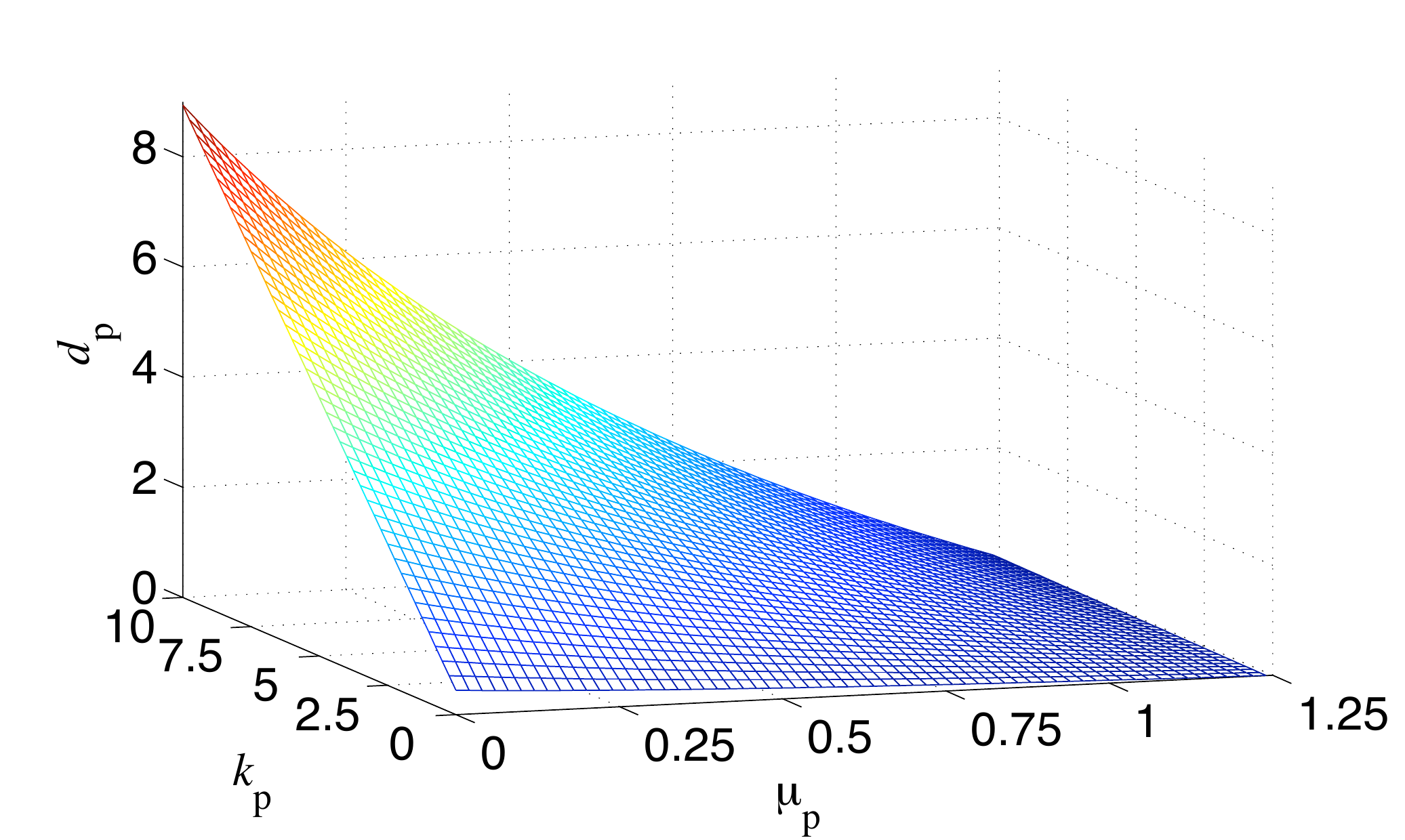}
\end{center}
\caption{Surface plot of the parameter space that satisfies the
conditions of Sec. III.A.1 for the mean run length and first passage
time. While both $\kp$ and $\drmp$ can have values along the positive
real line, $\mup$ has an upperbound.  If $\mup > 1.25 s^{-1}$ there
are no physically realistic values of $\kp$ and $\drmp$ that satisfy
the conditions $\langle X_{ap} \rangle=3.7$ and $T_{ap}=5.0$.  For
this plot, $v_a=0.88 \; \mu m$, $\mua=\drma=0.52 \; s^{-1}$, $k_a=
1.48 \;s^{-1}$.}
\label{fig:bestfitMT}
\end{figure}

\section{Analysis \& Discussion}

Having obtained analytical expressions for the mean run length and
duration, we can use experimental data from \cite{Ali08} to study
parameter assignment within our model. Because the experimental data
refer cooperative motor transport along microtubules and actin, we
will also divide the analysis into the two corresponding cases.

\subsection{Parameter Fitting}
\label{sec:parfit}

In the absence of external forces $v_a$ is constant, $v_p=0$, and for
fixed initial conditions, the model is characterized by 8
parameters. Only for a few of them it is possible to extract
estimates from the available literature (see Table
\ref{tab:parametervalues}). However, we can use certain biophysical
constraints stemming from \cite{Ali08} to reduce the parameter space
as much as possible. Throughout this subsection the analysis is 
performed with $\alpha=\beta =0$.\\

\noindent{\bf III.A.1 myoV-kinesin transport on MT.} Experimental
results from \cite{Ali08} show that the presence of a passive myoV
motor, in addition to an active kinesin motor increases the typical
run length of the cargo by two-fold and slightly decreases the
velocity by $\sim 15\%$. The same data show that the velocity and mean
run length of cargo in state $(1,0)$ and state $(1,1)$ are essentially
the same. Fig. \ref{fig:diffusiverun} is a graphical representation of
a possible cargo trajectory showing three processive and three
diffusive runs.  Since all processive runs are observed to occur with
the same velocity, the presence of the passive myosin does not affect
the drive of the active kinesin motor. Therefore, both states $(1,0)$
and $(1,1)$ are indistinguishable within each processive run.  These
observations suggest that we can assume $\drma = \mua$ in
Eq. \ref{eqn:master}. We used this assumption and the values of $v_a$,
$\mua$ from Table \ref{tab:parametervalues} to solve the system
consisting of Eq. \ref{eqn:Xapsimple} and Eq. \ref{eqn:Tapsimple} with the
constraints $\langle X_{ap} \rangle=3.7 \; \mu m$ and $T_{ap}=5.0\; s$
obtained from the experimental results in \cite{Ali08} for
microtubular transport.

The solution of this system leads to a specific value of $k_{a}=1.48
\; s^{-1}$, implying an average diffusion time of $1/\ka=0.68 \;s$,
consistent with the experimental results in \cite{Ali08}. In fact, the
average run length of a kinesin-myoV cargo complex on microtubules
lasts about 5 seconds and covers twice the distance of a Qdot with
only a single kinesin motor attached to it.  Therefore, the typical
cargo movement due to kinesin/myoV motor consists of two processive
steps (needing $\sim 2T_{a}=4.4 \; s$, see \cite{Ali08}) and one or
two diffusive ones before detachment since we assumed that the cargo
is initially in state $(1,1)$.  Given such observations, our obtained
value of $\ka$ is consistent with the experimental data. More
specifically, it suggests that on average there will be only one
single diffusive event in between two processive runs.

For the three remaining parameters we find that as long as $\mup <
1.25 s^{-1}$, we can always find $k_{\rm p}$ and $\drmp$ that satisfy the
physical constraints $\langle X_{ap} \rangle=3.7 \; \mu m$ and
$T_{ap}=5.0\; s$.  These results are shown in
Fig. \ref{fig:bestfitMT}.  The upper limit for $\mup$ implies an
average diffusion time of about $0.8$ seconds while in state $(0,1)$
and before detachment. However, the experimental results in
\cite{Ali07, Ali08} show an average diffusion time between $40$ and
$60$ seconds in the absence of kinesin.  The observed diffusion times
are consistent with the value for $k_a$ obtained above and suggest
that the detachment of the cargo from the microtubule is most likely
to happen while the motor is in state $(1,0)$.

Overall, these results suggest that myoV increases the processivity of
the cargo complex by keeping kinesin close enough to the track 
so that its reattach is accelerated. The tethering by myoV occurs 
without reduction in the intrinsic velocity of kinesin.\\

\begin{figure}[tb]
\begin{center}
\includegraphics[width=9cm]{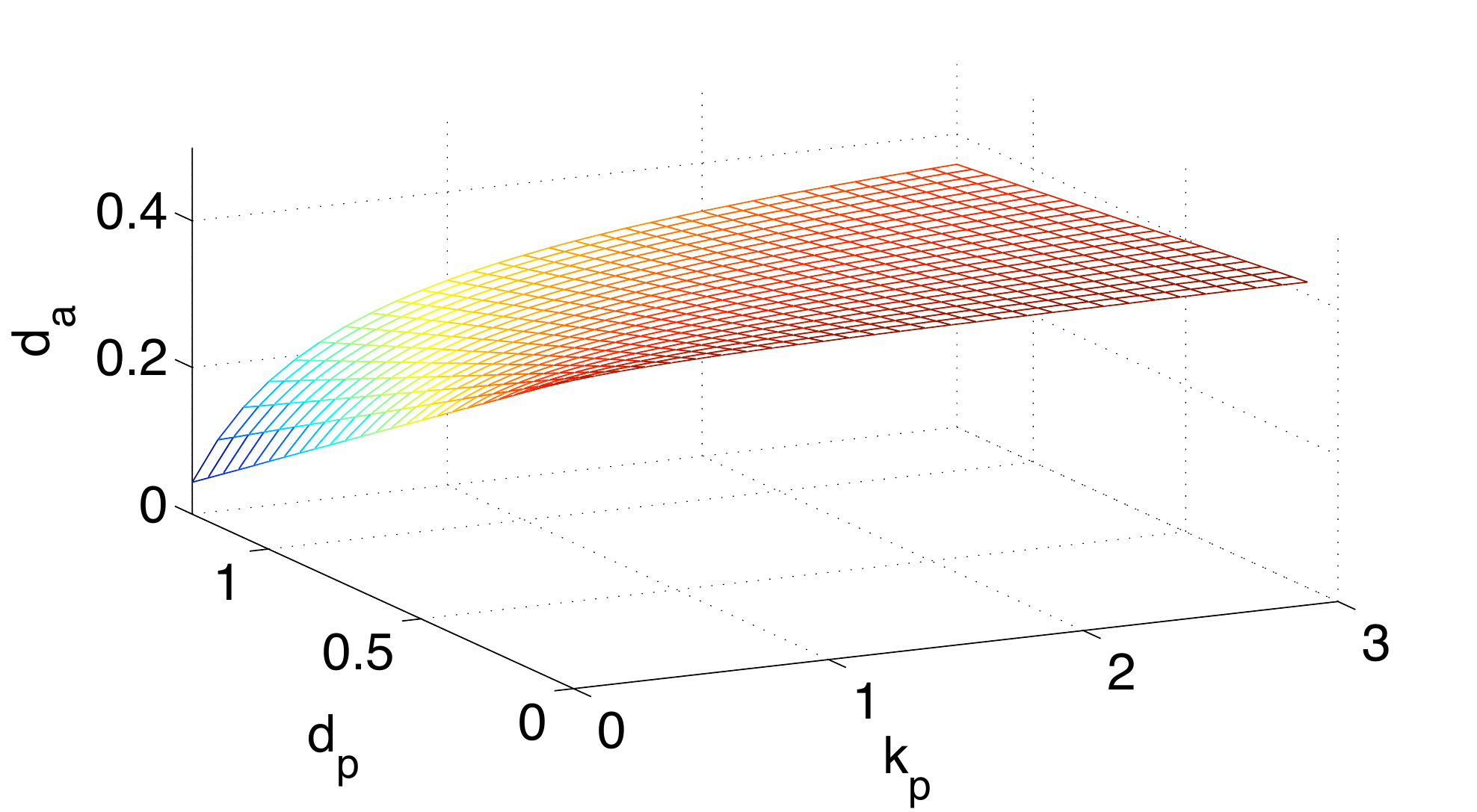}
\end{center}
\caption{Surface plot of the values of $\kp$, $\drmp$, and $\drma$
that satisfy the mean run length expression in Eq.  \ref{eqn:XapMyoV}
when, $v_a=0.46 \; \mu m$ and $\mua=0.60 \; s^{-1}$.  For $\kp \ll
\drmp$, $\drma$ tends to the value $0.42 s^{-1}$. This value is
equivalent to having a cargo complex consisting of only one active and
no passive motor.}
\label{fig:bestfitActin}
\end{figure}

\noindent{\bf III.A.2 myoV-kinesin transport on actin.} The most striking
finding observed using this experimental system is that the processivity 
is increased by the presence of a passive kinesin tether, but the average
velocity remains unchanged, suggesting that the passive tether 
helps keep the active myoV attached to the filament track
without affecting its velocity. In these experiments, one observes
longer, uninterrupted processive transport, with rare punctuated
moments of diffusive cargo motion. This suggests that the system is
predominantly in states $(1,j)$ and that once the active motor
detaches, the whole cargo system does too. These observations are
consistent with the structural/molecular attributes of this system,
since the electrostatic forces between actin and kinesin are too weak
to significantly reduce the myosin driven cargo velocity. Moreover,
since actin filaments are thin, once myosin detaches, the detachment
of kinesin is also fast. But if myosin holds kinesin proximal to the
actin filament, the kinesin attachment rate is also fast, since free
diffusion is hindered by the tether. Within this context 
$\drma \ne \mua$. We can also assume that $\ka=0$ and that $\mup
\rightarrow \infty$, leading to the following expressions for the mean 
run length and first passage time
\begin{equation}\label{eqn:XapMyoV}
\langle X_{ap} \rangle = \frac{(k_{\rm{p}}+
d_{\rm{p}}+\mu_{\rm{a}})v_{a}} 
{d_{\rm{a}}(k_{\rm{p}}+\mu_{\rm{a}}) + 
d_{\rm{p}}\mu_{\rm{a}}}, 
\langle T_{ap} \rangle = \frac{\drmp + \kp + \mua} 
{d_{\rm{a}}(k_{\rm{p}}+\mu_{\rm{a}}) + 
d_{\rm{p}}\mu_{\rm{a}}}.
\end{equation}
The above expressions are the same as Eq.\,\ref{eqn:Xa} and
Eq.\,\ref{eqn:Ta} for the particular choice of initial conditions that
we have used throughout this section.  Moreover, from Eq. \ref{eqn:Va}
we know that the velocity of this system is constantly equal to
$v_a$. Therefore, the model is able to predict the observed
experimental behavior.  We can now use $\langle X_{ap} \rangle=1.09 \;
\mu m$ and $T_{ap}=2.59 \; s$ (obtained from Table 1 in \cite{Ali08})
to plot the parameter space that satisfies either one of the
expressions in Eq. \ref{eqn:XapMyoV} but not both since they are
redundant under the assumption $\langle V_{ap} \rangle =v_a$.  Using
the expression for the mean run length in Eq. \ref{eqn:XapMyoV}, we
obtain the plot shown in Fig. \ref{fig:bestfitActin}. Here, we see
that for $\kp \gg \drmp$, $\drma$ reaches a maximum value of $d_a =
0.42 \; s^{-1}$. This quantity is smaller than $\mua$, a result that
confirms our interpretation that the increase in processivity is due
to the tethers ability to prevent detachment of the active motor.

From the properties of actin transport discussed
thus far, it seems natural to consider an ``effective'' detachment
rate $\mu_{\rm eff}$ that captures the dynamics of cargo transport in
this case:

\begin{equation}\label{eqn:effectivemu}
\mu_{\rm eff} =f(\mua,\mup,\kp,\drma,\drmp,v_a)=\frac{v_a}{\langle X_{ap} \rangle}.
\end{equation}
\noindent
From the results in \cite{Ali08}, we obtain $\mu_{\rm eff} \approx
0.42 \; s^{-1}$. This value is the same as the maximum value obtained
above, implying that the overall effect of kinesin is to prevent myoV
from detaching from the actin filament, without affecting the
intrinsic myoV transport velocity.

\begin{figure}[tb]
  \begin{center}
\includegraphics[width=9cm]{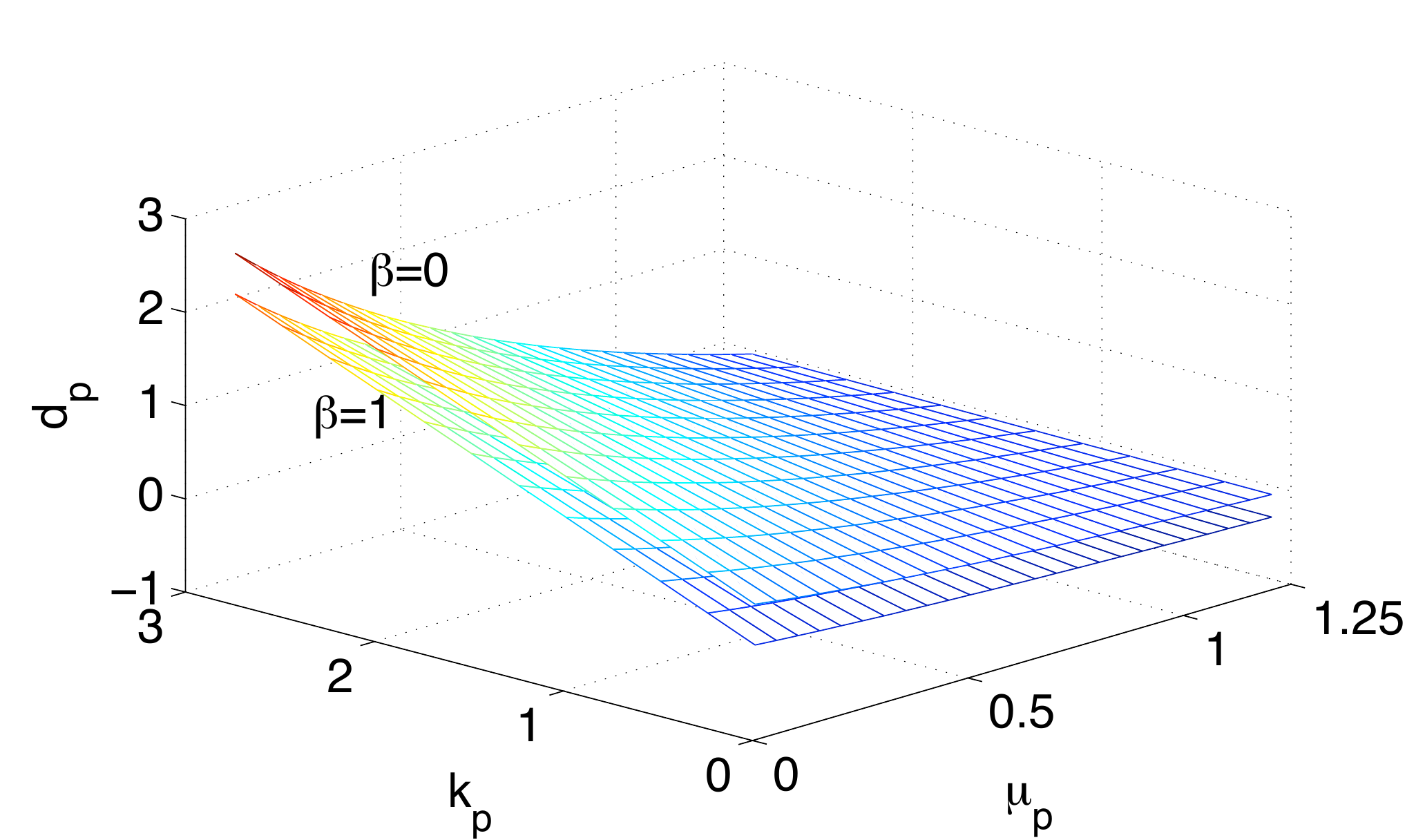}
  \end{center}
\caption{Plot of the dependence of $\kp$, $\drmp$, and $\mup$ on
$\beta$ for transport along microtubules. The values of the other
parameters are the same as the ones in Fig.
\ref{fig:bestfitMT}. Increasing the number of cargo complexes in state
$(1,0)$ at time $t=0$ leads to smaller values of tether detachment
from state $(1,1)$ for all values of $\kp$ and $\mup$.}
\label{fig:iconMT}
\end{figure}

\subsection{Dependence on Initial Conditions}

We now investigate how the model fits data as $\alpha$ and $\beta$ in
Eq.  \ref{eqn:metransform} vary between zero and one.  Based on the
experimental methodology of \cite{Ali08}, we expect both $\alpha$ and
$\beta$ to be small.  In fact, only Qdots that show a displacement
greater than $0.3 \; \mu m$ where included in the data, with shorter
trajectories discarded, essentially reducing $\alpha$ to zero.  The
argument for $\beta$ being small also relies upon the experimental
methods of Ali {\it et al.}. To ensure that each Qdot had at most one
active motor and at least a passive one attached to it, they prepared
a solution where passive motors were in excess, with molar ratio 16:1
\cite{Ali08}.\\

\noindent{\bf III.B.1 myoV-kinesin transport on MT.} As discussed above,
in the case of microtubular transport, we can use the approximation
$\drma=\mua$. Unlike the case with simple initial conditions ($\alpha
= \beta = 0$), including this constraint in Eqs.\,\ref{eqn:Xap} and
\ref{eqn:Tap} does not yield a simple analytical solution for the mean
run length and first passage time before detachment. Therefore, we
performed a numerical study of the dependence on the initial
conditions and found that for every choice of $\alpha$ and $\beta$
there is only one value of $k_a$ that satisfies the experimental
results $\langle X_{ap} \rangle=3.7 \; \mu m$ and $T_{ap}=5\; s$. In
addition, this value depends only on the initial fraction of cargoes
in state $(0,1)$ and not on $\beta$. In particular, $\ka$ is a linear
function of $\alpha$, with slope $m=1.26\; s^{-1}$, giving us a range
of predicted values from $\ka = 1.48\; s^{-1}$ (if $\alpha = 0$) to
$\ka = 2.74\; s^{-1}$ (if $\alpha = 1$).  The increase of predicted
$\ka$ with $\alpha$ is not surprising since the higher the probability
the systems starts in state $(0,1)$, the faster the kinesin motor will
have to bind to the microtubule to satisfy the given time
constraint. Conversely, the values of the other free parameters in the
model ({\it i.e.} $\drmp$, $\mup$, and $\kp$) depend only on the value
of $\beta$. This dependence is shown in Fig. \ref{fig:iconMT} for the
limit cases $\beta=0$ and $\beta=1$ (in both cases $\alpha=0$). From
this plot we notice how an increase in the percentage of cargo
complexes in state $(0,1)$ at $t=0$, shifts the parameter surface down
along the $\drmp$ axis. As a result the parameter space itself in
Fig. \ref{fig:iconMT} is reduced, since any combination of parameters
resulting in $\drmp<0$ is unphysical. From this analysis it seems that
all possible initial conditions can explain the experimental
data. None of the qualitative observations made in Sec. III.A.1 would
change, unless $\alpha > 0$. In this case, the average cargo movement
would consist of two diffusive steps, one at the beginning of the
motion and one at the end of the first processive step.\\

\begin{figure}[tb]
  \begin{center}
 \includegraphics[width=9cm]{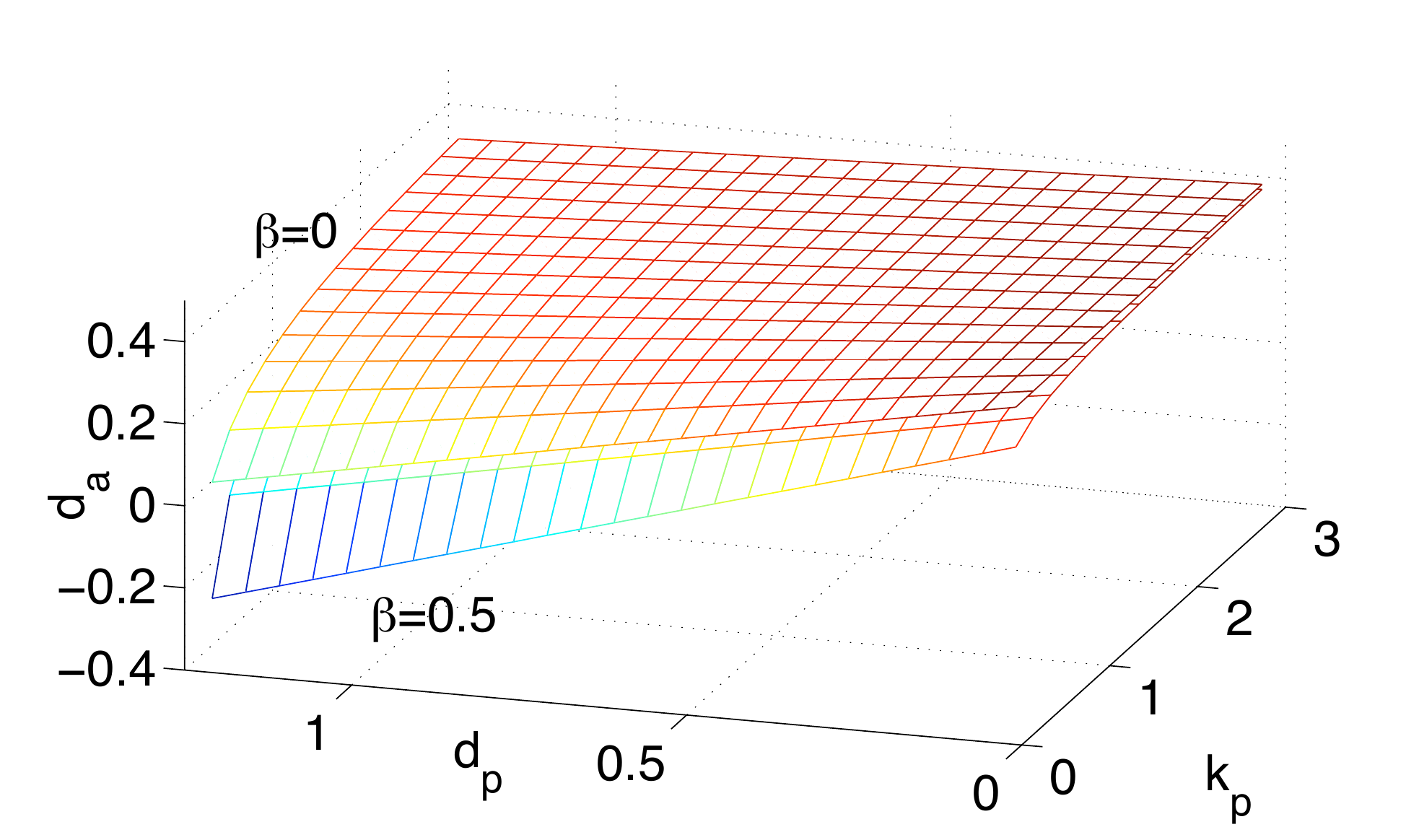}
  \end{center}
\caption{Plot of the dependence of $\kp$, $\drmp$, and $\drma$ on
$\beta$ for transport along actin. The values of the other parameters
are the same as the ones in Fig.  \ref{fig:bestfitActin}. Increasing
the number of cargo complexes in state $(1,0)$ at time $t=0$ leads to
smaller values of active motor detachment rate from state $(1,1)$.
Moreover, some of the values of $\kp$ and $\drmp$ lead to negative
values of $\drma$ although the limiting behavior for $\kp \ll \drmp$
is still the same as for the case $\beta=0$.}
\label{fig:iconActin}
\end{figure}

\noindent{\bf III.B.2 myoV-kinesin transport on actin.} 
Using the same assumptions discussed in Section III.A.2, 
we find the following simplified expression,

\begin{equation}\label{eqn:Xapic}
\langle X_{ap} \rangle = v_a\frac{(\beta
(\drma-\mua)+(1-\alpha)(\kp+\mua))+\drmp(1-\alpha)}
{\drma (\kp+\mua)+\drmp + \mua},
\end{equation}

\noindent
indicating a mean run length that decreases linearly as $\alpha$
increases. This is physically expected since we assumed $\ka=0$, which
implies that diffusional states are not allowed to transition to the
processive (1,1) state.  How $\langle X_{ap} \rangle$ above scales
with $\beta$ depends instead on the difference between $\mua$ and
$\drma$. The detachment rate of myoV from actin in state $(1,0)$ is
about $0.6 \; s^{-1}$, and we see from Fig.\,\ref{fig:bestfitActin}
that this value is never reached (we also verified this result
asymptotically). Then, the mean run length of the cargo complex is a
decreasing function of the probability of being in state $(1,0)$ at
time $t=0$. 

The overall effect of a higher value of $\beta$ is to lower the
best-fit value of $\drma$ and to reduce the range of physically
meaningful parameters, as shown in Fig. \ref{fig:iconActin}.  Similar
to what we mentioned in Section III.B.1, the experimental data can be
fit to all possible initial conditions.  However, if $\drmp$ is too
small, the condition for the mean run length from \cite{Ali08} cannot
be satisfied if $\beta$ is too large. This is reflected by the
negative values of the $\beta = 0.5$ fit to $d_{\rm a}$ in
Fig. \ref{fig:iconActin}.  These results reinforce the notion that the
increased processivity of myoV due to the presence of kinesin depends
on the latter's ability to keep the active motor attached to the track
for a longer period of time.

\subsection{Sensitivity Analysis}
\label{sec:sa}

We conclude this section by performing both local and global
sensitivity analysis of our model output on the model
parameters. Since our goal is to determine the effect of cooperation
among different molecular motors on the processivity of cargo
transport, we select $\langle X_{ap} \rangle$ ({\it e.g.} the mean run
length before detachment) as the output of interest. 

The simplest local sensitivity analysis evaluates the partial
derivatives of the mean run length before detachment with respect to
each of the unknown parameters ({\it e.g.} $\partial \langle X_{ap}
\rangle / \partial \drmp$). This determines how sensitive the output
is to quantitative changes in each of the kinetic parameters
\cite{Saltelli05}.

Since local sensitivity analysis is best suited to evaluating output
that is linear in the parameters, we will also consider global
sensitivity analysis on $\langle X_{ap} \rangle$.  This analysis will
determine which among the unknown parameters of the model would be
most responsible for experimental variation of the output
\cite{Saltelli04}. This analysis is global in the sense that it spans
all of the parameter space.  It is model-free, and gives the same
result as local sensitivity analysis if the analyzed models are linear
\cite{Saltelli04}.  Let us write $\Omega=\left\{ \mup, \mua, \drmp,
\drma, \kp, \ka, v_a \right\}$ as the input space, and $\Omega_{i}$,
$i=1,2,\dots,7$ as $i$-th input in $\Omega$. Then we can define the
first-order sensitivity for a fixed input $\Omega_i$ as:
\begin{equation}\label{eqn:si}
S_{i} \equiv \frac{V[ E(\langle X_{ap} \rangle | \Omega_i)]}{V[\langle X_{ap} 
\rangle]},
\end{equation}
where $E(\langle X_{ap} \rangle| \Omega_i)$ is the expected value of
the mean run length obtained by uniformly sampling over all other
parameters $\Omega_{j \ne i}$, $V[ E(\langle X_{ap} \rangle |
\Omega_i)]$ is the variance of the expected mean run length over the
parameter $\Omega_i$, and $V[\langle X_{ap} \rangle]$ is the
unconditional variance of the mean run length. The parameter with
highest first order sensitivity index is the one which most influences
the variation of the mean run length according to the global
sensitivity analysis approach. Global sensitivity analysis can also be
used to assess the joint effect of more than one input.  We define the
second order sensitivity (also known as two-way interaction) as
\begin{figure}[tb]
  \begin{center}
 \includegraphics[width=9cm]{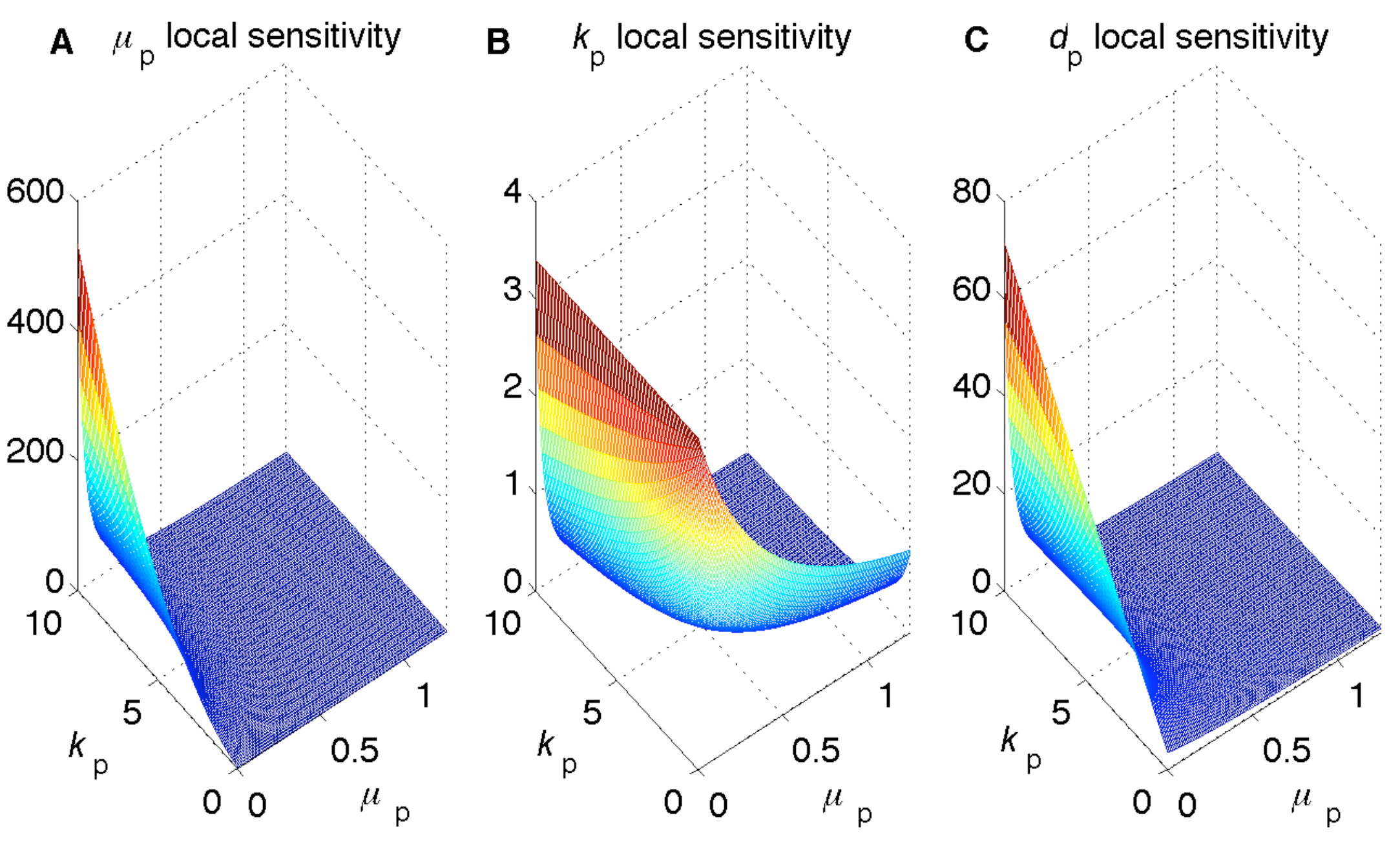}
  \end{center}
\caption{Local sensitivity analysis of free model parameters in the case of 
kinesin/myoV cargo transport on microtubules.
The parameters for all  three of these graphs are $v_a=0.88 \; \mu m/s$, 
$\drma=\mua=0.52 \; s^{-1}$, $\ka=1.48 \; s^{-1}$, and $\drmp=0.5 \; s^{-1}$. 
 {\bf A}: Sensitivity analysis with respect to $\mup$, in this case 
 the z-axis represents $\partial \langle X_{ap} \rangle / \partial \mup$. 
 {\bf B}: Sensitivity analysis with respect to $\kp$, in this case 
 the z-axis represents $\partial \langle X_{ap} \rangle / \partial \kp$ 
 {\bf C} Sensitivity analysis with respect to $\drmp$, in this case 
 the z-axis represents $\partial \langle X_{ap} \rangle / \partial \drmp$.}
\label{fig:sensitivityMT}
\end{figure}

\begin{equation}\label{eqn:si2}
S_{ij} \equiv \frac{V[ E(\langle X_{ap} \rangle | \Omega_i, \Omega_j)]}
{V[\langle X_{ap} \rangle]},
\end{equation}
where $E(\langle X_{ap} \rangle | \Omega_i , \Omega_j)$ is the
expected value of the mean run length given fixed values of $\Omega_i$
and $\Omega_j$. Higher order global sensitivity indexes can be
analogously defined. We apply local and first and second order global
sensitivity analyses to both experimental cases:\\

\noindent{\bf III.C.1 myoV-kinesin transport on MT.}  Representative
results of the local sensitivity analysis for the microtubule case are
plotted in Fig. \ref{fig:sensitivityMT}.  Under certain regimes,
$\mup$ has the greatest influence on the mean run length before
detachment, followed by $\drmp$, with $\kp$ as the least influential
among the three parameters.  To further investigate the results from
the local sensitivity analysis, we determine the first and second
order global sensitivity indexes for all free parameters. These
results are listed in the first row of Table \ref{tab:saresults}. We
find that the parameter that is responsible for most of the variation
in $\langle X_{ap} \rangle$ is $\drmp$.

Both of the analyses predict that $\kp$ has the least influence on the
mean run length, but they differ in their ranking of $\mup$ and
$\drmp$. This difference arises from the nonlinearity of
Eq. \ref{eqn:Xap} and the intrinsic differences among the two types
analyses.  Local sensitivity analysis suggests that if we could
control the values of the parameters of the system, we would effect
the largest changes in $\langle X_{ap} \rangle$ by altering the rate
of detachment of myoV while in state $(0,1)$. If experimentally, we
are sampling parameter space, global sensitivity analysis predicts
that by correctly determining $\drmp$ we can achieve the most
reduction in the variability of the mean run length. \\

\begin{table}[tb]
\begin{center}
\caption{First and second order sensitivity indexes for microtubule and 
actin cargo transport. To determine these indexes we sampled $\drmp$ 
and $\kp$ uniformly in $[0,20]$ with step $0.05$. For the microtubule case 
we sample $\mup$ in $[0,1.25]$ with step $0.01$. For the actin case 
we sampled $\drma$ in $[0,0.42]$ with step $0.01$.}
\label{tab:saresults}
\begin{tabular}{lccccccccr}
\hline
 & $S_{\kp}$ & $S_{\drmp}$ & $S_{\mup}$ & $S_{\drma}$
& $S_{\kp,\mup}$ & $S_{\kp,\drmp}$ & $S_{\drmp,\mup}$ & 
$S_{\drma, \kp}$ & $S_{\drma,\drmp}$\\
\hline \\
MT & 0.06 & 0.27 & 0.1 & -- & 0.20 & 0.35 & 0.82 & -- & -- \\
Actin & 0.1 & 0.33 & -- & 0.23 & -- & 0.45 & -- & 0.37 & 0.81 \\
\hline
\end{tabular}
\end{center}
\end{table}

\begin{figure}[tb]
  \begin{center}
 \includegraphics[width=9cm]{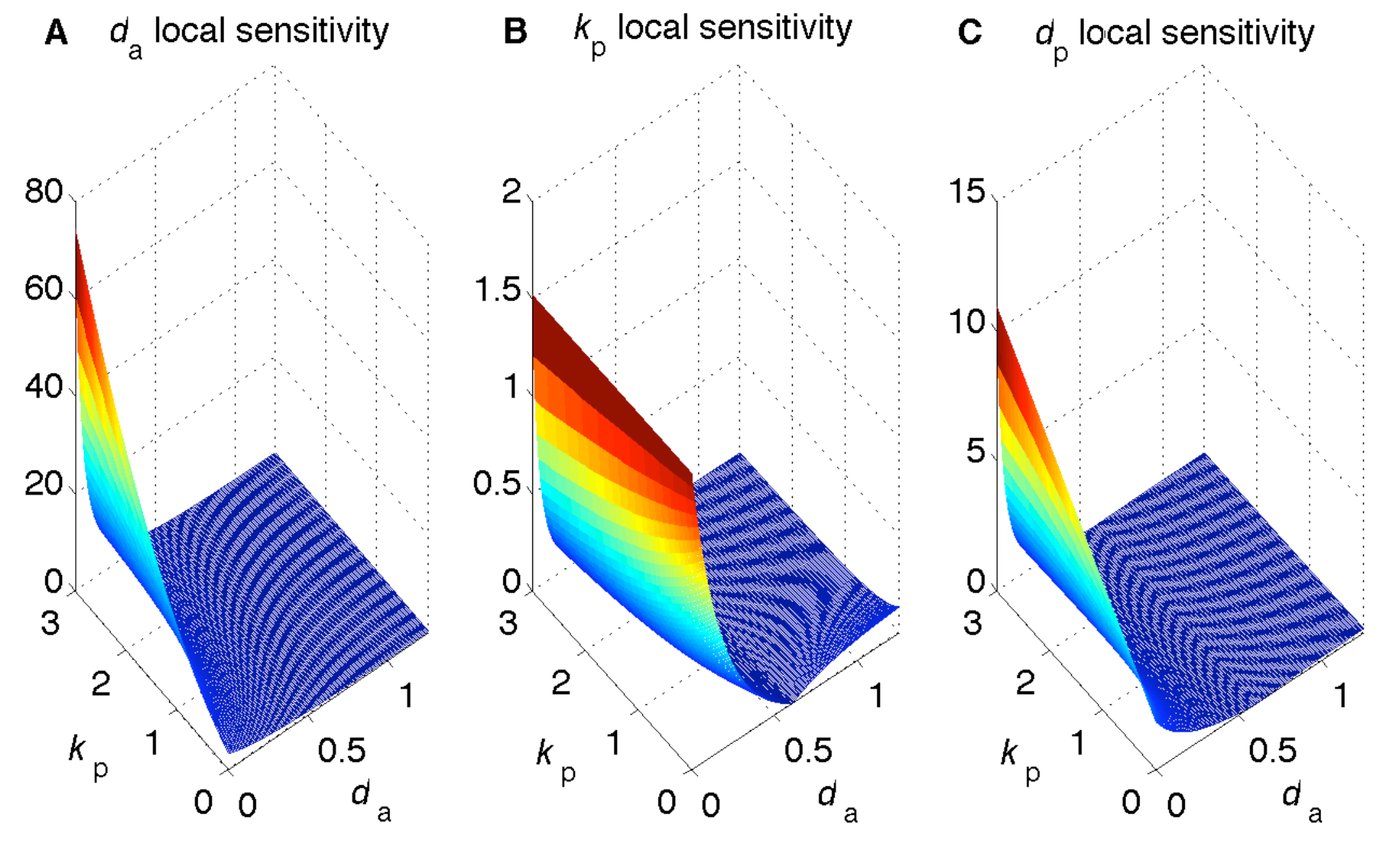}
  \end{center}
\caption{Local sensitivity analysis of free model parameters in the 
case of myoV/kinesin cargo transport on actin. The parameters for all  three of 
these graphs are $v_a=0.46 \; \mu m/s$, $\mua=0.6 \; s^{-1}$, 
$\ka=0 \; s^{-1}$, $\mup \ll 1$ and $\drmp=0.5 \; s^{-1}$. 
{\bf A}: Sensitivity analysis with respect to $\mup$, in this case 
 the z-axis represents $\partial \langle X_{ap} \rangle / \partial \drma$. 
 {\bf B}: Sensitivity analysis with respect to $\kp$, in this case 
 the z-axis represents $\partial \langle X_{ap} \rangle / \partial \kp$ 
 {\bf C} Sensitivity analysis with respect to $\drmp$, in this case 
 the z-axis represents $\partial \langle X_{ap} \rangle / \partial \drmp$.}
\label{fig:sensitivityActin}
\end{figure}

\noindent{\bf III.C.2 myoV-kinesin transport on actin.}  The local and
global sensitivity analyses also give different results in the case of
cargo transport along actin filaments.  Three representative plots of
the local sensitivity analysis are shown in
Fig. \ref{fig:sensitivityActin}. These show that the detachment rate
of myoV from actin when both motors are attached is the most
influential parameter with respect to the mean run length before
detachment.  However, the first order sensitivity index of $\drma$ is
about $1/3$ smaller than the same index for $\drmp$ (see Table
\ref{tab:saresults}), making the detachment rate of kinesin when in
state $(1,1)$ the parameter more responsible for the variance in
$\langle X_{ap} \rangle$. In this case, the sensitivity to $\drmp$ is
consistent with the role of the passive tether in preventing myoV
detachment.


Interestingly, both cases (actin and microtubule) have $\drmp$ as the
free parameter with highest first order sensitivity and $\kp$ as the
lowest one. Moreover, if we want to reduce by at least $80\%$ the
uncertainty in the determination of the mean run length, we can do so
by exactly determining $\drmp$ and $\mup$ for the microtubule case and
$\drmp$ and $\drma$ for the actin case (cf. Table
\ref{tab:saresults}).

\section{Summary \& Conclusions}

We presented a stochastic model that describes the cooperative
behavior between two different motors attaching cargoes to a
cytoskeletal filament. Of the two motors, one acts as engine, moving
the cargo unidirectionally along the filament, while the other acts as
a tether. We applied the model to the data from \cite{Ali08}, where
the authors studied {\it in vitro} the cooperative behavior of kinesin
and myosin V when moving cargo along microtubules and actin.
Experimental visualization indicates significant diffusive dynamics
for cargo transport along microtubules in the presence of myosin V,
while cargoes transported along actin did not exhibit diffusive
dynamics. A consistent interpretation of these observations is that
along microtubules, tethers (myosin) predominantly enhances
reattachment of the active motor (kinesin).  In the case of transport
along actin, the kinesin tether acts to prevent detachment of the
active myosin motor. This interpretation has been verified within our
model. In fact, for the case of microtubule transport we found that
the reattachment rate ($\ka$) of kinesin to the filament when the
tether is attached is three times faster than its corresponding
detachment rate. For the case of actin, we found that the detachment
rate of myoV when kinesin is attached to the filament is slower than
in the absence of the tether for all values in the explored parameter
space.

Although most of the modeling effort described in the literature use a
very detailed representation of the physical and chemical properties
of the molecular motor under investigation, such analysis becomes much
more complex when considering multiple cooperating motors.  Coarser
models such as ours can be straightforwardly extended to incorporate
cargo systems with $M$ active motors and $N$ tethers. The state of
such cargo complex will be characterized by a $M+N$-tuple, whose first
$M$ components have value $1$ or $0$ depending on whether the active
motors they represent are attached to the cytoskeletal filament or
not. Conversely, the last $N$ components will have value $1$ or $0$
depending on the attachment status of each passive motor. Proceeding
as we did in Section \ref{sec:model}, we can define a Master equation
from which mean run lengths and first passage times to full
detachment may be determined. As done in the present case, such a
model could be used to find the optimal configuration of motor and
tethers that fits the available experimental data. 


\section{Acknowledgments} 
\noindent
The authors are grateful to Y. Ali, H. Lu, and D. Warshaw for their
clarifying comments.  This work was supported by grants from the NSF
(DMS-0349195, DMS-0719462) and the NIH (K25 AI41935).


\begin{thebibliography}{99}


\bibitem{Vale03} R. D. Vale,
{\it Cell}, 2003, {\bf 112}, 467.

\bibitem{Mallik06} R. Mallik and S.P. Gross,
{\it Physica A}, 2006, {\bf 372}, 65.

\bibitem{Lodish05} H. Lodish, A. Berk, P. Matsudaira, C.A. Kaiser, M. Krieger, M.P.
Scott, S.L. Zipursky and J. Darnell,
{\it W.H. Freeman Co}, 2005, {\bf 5th Ed.}

\bibitem{Gross07review} S.P. Gross,
{\it Curr. Biol.}, 2007, {\bf 17}, R277.

\bibitem{Welte04} M.A. Welte,
{\it Curr. Biol.}, 2004, {\bf 14}, R525.

\bibitem{Metha99} A.D. Metha, R.S. Rock, M. Rief, J.A. Spudich, M.S. Mooseker and R.E. Cheney,
{\it Nature}, 1999, {\bf 400}, 590.

\bibitem{Gross07} S.P. Gross, M. Vershinin and G.T. Shubeita,
{\it Curr. Biol.}, 2007, {\bf 17}, R478.

\bibitem{Block90} S.M. Block, L.S.B. Goldstein and B.J. Schnapp,
{\it Nature}, 1990, {\bf 348}, 348.

\bibitem{Visscher99} K. Visscher, M.J. Schnitzer and S.M. Block,
{\it Nature}, 1999, {\bf 400}, 184.

\bibitem{Shiroguchi07} K. Shiroguchi and K. Kinosita Jr.,
{\it Science}, 2007, {\bf 316}, 1208.

\bibitem{Rief00} M. Rief, R.S. Rock, A.D. Mehta, M.S. Mooseker, R.E. Cheney and J.A. Spudich,
{\it Proc. Natl. Acad. Sci. USA}, 2000, {\bf 97}, 9482.

\bibitem{Carter05} N.J. Carter and R.A. Cross,
{\it Nature}, 2005, {\bf 435}, 308.

\bibitem{Svoboda94} K. Svoboda and S.M. Block,
{\it Cell}, 1994, {\bf 77}, 773.

\bibitem{Kawaguchi00} K. Kawaguchi and S. Ishiwata,
{\it Biochem. Biophys. Res. Comm.}, 2000, {\bf 272}, 895.

\bibitem{Clemen05} A.E. Clemen, M. Vilfan, J. Jaud, J. Zhang, M. Barmann and M. Rief,
{\it Biophysical J.}, 2005, {\bf 88}, 4402.

\bibitem{Fisher99} M.E. Fisher and A.B. Kolomeisky,
{\it Physica A}, 1999, {\bf 274}, 241.

\bibitem{Kolomeisky03} A.B. Kolomeisky and M.E. Fisher,
{\it Biophysical J.}, 2003, {\bf 84}, 1642.

\bibitem{Skau06} K. I. Skau, R. B. Hoyle and M. S. Turner,
{\it Biophysical J.}, 2006, {\bf 91}, 2475.

\bibitem{Kolomeisky07} A.B. Kolomeisky, and M.E. Fisher,
{\it Annu. Rev. Phys. Chem.}, 2007, {\bf 58}, 675.

\bibitem{Vilfan08} A. Vilfan,
{\it Biophysical J.}, 2008, {\bf 94}, 3405.

\bibitem{Kunwar08} A. Kunwar, M. Vershinin, J. Xu and S.P. Gross
{\it Curr. Biol.}, 2008, {\bf 18}, 1.

\bibitem{Ali07} M. Y. Ali, E.B. Krementsova, G.G. Kennedy, R. Mahaffy, T.D. Pollard, K. M. Trybus and D. M. Warshaw,
{\it Proc. Natl. Acad. Sci. USA}, 2007, {\bf 104}, 4332.

\bibitem{Ali08} M. Y. Ali, H. Lu, C. S. Bookwalter, D. M. Warshaw and K. M. Trybus,
{\it Proc. Natl. Acad. Sci. USA}, 2008, {\bf 105}, 4691.

\bibitem{Alivisatos05} A.P. Alivisatos, W. Gu and C. Larabell,
{\it Annu. Rev. Biomed. Eng.}, 2005, {\bf 7}, 55.

\bibitem{Milescu06} L. S. Milescu, A. Yildiz, P. R. Selvin and F. Sachs,
{\it Biophysical J.}, 2006, {\bf 91}, 3135.

\bibitem{Klumpp05} S. Klumpp and R. Lipowsky,
{\it Proc. Natl. Acad. Sci. USA}, 2005, {\bf 102}, 17284.

\bibitem{Saltelli05} A. Saltelli, M. Ratto, S. Tarantola and F. Campolongo,
{\it Chem. Rev.}, 2005, {\bf 105}, 2811.

\bibitem{Saltelli04} A. Saltelli, S. Tarantola, F. Campolongo and M. Ratto,
{\it W.H. Freeman Co}, 2005, {\bf 5th Ed.}

\end{thebibliography}
\end{document}